\documentclass[prl,
               preprint,
               superscriptaddress,
               onecolumn,
               showpacs,
               longbibliography,
               ]{revtex4-1}
               
\usepackage{graphicx}   
\usepackage{dcolumn}    
\usepackage{bm}         
\usepackage{amsmath}
\usepackage{amssymb}
\usepackage[usenames]{color}      

\begin{document}

\title{Vertically bounded double diffusive convection in the fingering regime: comparing no-slip vs free-slip boundary conditions}

\author{Yantao Yang}
\affiliation{Physics of Fluids Group, MESA+ Research Institute, and J. M. Burgers Centre for Fluid Dynamics, University of Twente, PO Box 217, 7500 AE Enschede, The Netherlands.}

\author{Roberto Verzicco}
\affiliation{Physics of Fluids Group, MESA+ Research Institute, and J. M. Burgers Centre for Fluid Dynamics, University of Twente, PO Box 217, 7500 AE Enschede, The Netherlands.}
\affiliation{Dipartimento di Ingegneria Industriale, University of Rome ``Tor Vergata'', Via del Politecnico 1, Roma 00133, Italy}
                    
\author{Detlef Lohse}
\affiliation{Physics of Fluids Group, MESA+ Research Institute, and J. M. Burgers Centre for Fluid Dynamics, University of Twente, PO Box 217, 7500 AE Enschede, The Netherlands.}
\affiliation{Max-Planck Institute for Dynamics and Self-Organization, Am Fassberg 17, 37077 G\"{o}ttingen, Germany.}

\date{\today}

\begin{abstract}
Vertically bounded fingering double diffusive convection (DDC) is numerically investigated, focusing on the influences of different velocity boundary conditions, i.e. the no-slip condition which is inevitable in the lab-scale experimental research, and the free-slip condition which is an approximation for the interfaces in many natural environments, such as the oceans. For both boundary conditions the flow is dominated by fingers and the global responses follow the same scaling laws, with enhanced prefactors for the free-slip cases. Therefore, the laboratory experiments with the no-slip boundaries serve as a good model for the finger layers in the ocean. Moreover, in the free-slip case although the tangential shear stress is eliminated at the boundaries, the local dissipation rate in the near-wall region may exceed the value found in the no-slip cases, which is caused by the stronger vertical motions of fingers and sheet structures near the free-slip boundaries. This counter intuitive result might be relevant for properly estimating and modelling the mixing and entrainment phenomena at free-surfaces and interfaces.
\end{abstract}


\maketitle

Double diffusive convection (DDC) -- the convection flow of two scalars (e.g. temperature and concentration) which affect the density -- is ubiquitous in many natural environments. The terrestrial system of the greatest relevance is the oceanic flow~\citep{Turner1985,Schmitt1994,Schmitt2003,Schmitt2005}, where the density of seawater mainly depends on temperature and salinity. Originally proposed as an oceanographical curiosity~\citep{Stommel1956}, DDC has drawn lots of attention since it plays an important role in the oceanic mixing, e.g.~see a comprehensive review in the recent book of~\citet{Radko2013} and the references therein. A particularly fascinating phenomenon of DDC flows are the salt fingers, which occur when a fluid layer experiences an unstable salinity gradient and a stable temperature gradient. Salt fingers can even grow when the overall stratification is stable~\citep{Stern1960}. Conditions favouring salt fingers are present in most subtropic oceans~\citep{You2002}. Fingering DDC can induce intense vertical mixing~\citep{Lee_etal:GRL:2014} and may even attenuate the oceanic signatures of climate changes~\citep{Johnson_Kearney:GRL:2009}. 

Most early experiments on finger convection employed a sharp interface from which the salt fingers grow and extend freely in the vertical direction~\cite{Schmitt2003}. When starting from a thick region with both temperature and salinity gradients, one finger layer or a stack of alternating convection and finger layers may develop, depending on the control parameters~\citep{Linden1978,Krishnamurti2003,Krishnamurti2009}. Three-dimensional (3D) direct numerical simulations (DNS) were also conducted and provided detailed informations on DDC, such as simulations in a fully periodic box with uniform background gradients~\citep{Rosenblum2011,Stellmach2011,Traxler2011,Radko2014} and those bounded by two parallel plates~\citep{Paparella2012}. In the recent experiments employing electrodeposition cells by Tilgner and coworkers~\citep{Hage_Tilgner2010,Kellner_Tilgner2014}, one single finger layer was observed between top and bottom boundaries for both stable and unstable stratification. Those experiments provide a good platform to investigate the vertical scalar transport and flow structures of finger layer. Our previous DNS successfully reproduced most key observations in experiments and good agreement was obtained between our numerical results and the experimental results~\citep{ddcjfm2015,ddcpnas16}.

However, an inevitable difference between the experiments and the oceanic finger layers is that the experiments was done with no-slip boundaries which do not exist in the ocean. Therefore the relevance of these experiments for oceanic DDC flow has been questioned. To clarify the relevance of this difference, in this paper by using DNS, we carry out a comparison between finger layers bounded by two no-slip plates which are the same as the experiments (e.g.~\citep{Hage_Tilgner2010,Kellner_Tilgner2014}) and those bounded by two free-slip plates which model the oceanic finger layers usually bounded by two homogeneous convection layers as in the thermohaline staircase. Similar studies were conducted for (rotating) RB flow~\citep{Petschel2013,Ostilla2014}. The current study is required to apply the experimental results of DDC flow to oceanic flow~\citep{Kellner_Tilgner2014}. Moreover, the comparison between different boundary conditions reveals some surprising characteristics of fingering DDC flow. 

Consider DDC flow between two parallel plates which are perpendicular to the direction of gravity and separated by a height $L$. At the two plates both temperature and salinity are kept constant. The two Prandtl numbers, i.e.~the ratio between the scalar molecular diffusivity to viscosity, are fixed at $Pr_T=7$ and $Pr_S=700$, which are the typical values of seawater. We confine ourself in the finger regime with the top plate having both higher temperature and salinity. The flow is driven by the salinity difference $\Delta_S$ between two plates and stabilised by the temperature difference $\Delta_T$. The strength of the buoyancy force associated with the scalar field $\zeta=T$ or $S$ is measured by the Rayleigh number $Ra_\zeta = (g \beta_\zeta L^3 |\Delta_\zeta|)/(\lambda_\zeta \nu)$, with $g$ being the gravitational acceleration and $\beta_\zeta$ the positive expansion coefficient, respectively. The density ratio, which reflects the relative strength of the buoyancy force induced by temperature difference to that by salinity difference, can then be calculated as $\Lambda=(\beta_T\Delta_T)/(\beta_S\Delta_S) = (Pr_S Ra_T) / (Pr_T Ra_S)$. 

The flow quantities are nondimensionalized by $L$, $|\Delta_T|$, $|\Delta_S|$, and the free fall velocity $U=\sqrt{g \beta_S |\Delta_S| L}$. We numerically solve the incompressible Navier-Stokes equation within the Oberbeck-Boussinesq approximation for the velocity $u_i$ with $i=1,2,3$, pressure $p$, and nondimensionalized temperature $\theta$ and salinity $s$, respectively. The subscript ``3'' denotes the component in the vertical direction opposite to gravity. A reliable finite-difference code was utilised together with a multiple resolution technique in order to cope with the very different diffusivities of the two scalars~\citep{ddcjfm2015,multigrid2015}. At the two plates either no-slip or free-slip boundary conditions are imposed for the tangential velocity components and the non-penetration condition for the normal velocity component, respectively. In horizontal directions we apply periodic boundary conditions. The aspect ratio is chosen such that the horizontal size of the domain is much larger than the horizontal scale of the flow structures, i.e. the salt fingers. 

\begin{figure*}[h!]
\begin{center}
\noindent\includegraphics[width=\textwidth]{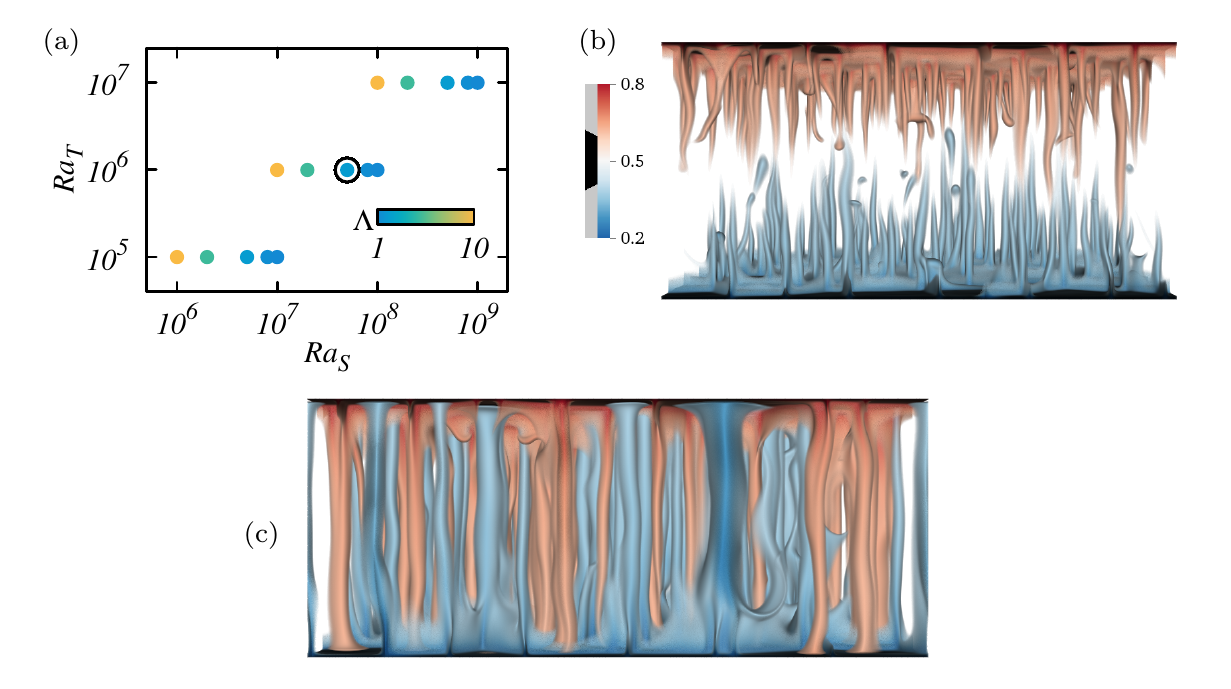}%
\caption{(a) The explored parameters shown in the $Ra_S-Ra_T$ plane and coloured by the density ratio $\Lambda$. (b, c) The volume rendering of salt fingers for $(Ra_S, Ra_T)=(5\times10^7, 1\times10^6)$ (marked by a black circle in panel a) with (b) the no-slip boundary condition and aspect ratio $2.0$, and (c) the free-slip boundary condition and aspect ratio $2.4$, respectively. The two plots share exactly the same colormap and opacity settings.}
\label{fig:finger}
\end{center}
\end{figure*}

Three different values of $Ra_T$ are considered, namely~$Ra_T=10^5$, $10^6$, and $10^7$. For each $Ra_T$ five cases are simulated with $\Lambda$ ranging from $1$ to $10$, i.e. in the salt finger regime. The explored phase space is shown in Fig.~\ref{fig:finger}a. For every set of control parameters two simulations were conducted with either no-slip or free-slip boundary condition. Initially the fluid is at rest, the temperature field has a vertically linear distribution, and the salinity field is uniform and equal to the mean of the values at the top and bottom plates, respectively. From this initial condition one single finger layer develops between two plates in all simulations. Figs.~\ref{fig:finger}b and c display the three-dimensional volume rendering of salt fingers for $(Ra_S, \Lambda)=(5\times10^7, 2.0)$ with different boundary conditions. The colormap and opacity settings are exactly the same in the two plots. The flow morphology is essentially the same for the two boundary conditions: The vertically oriented salt fingers occupy the whole bulk region, while near the two plates sheet-like structures connect the roots of the adjacent fingers. Usually fingers are associated with slender convection cells. Clearly, the salt fingers in the free-slip case are much stronger than those in the no-slip case: This is expected since the no-slip boundaries not only exert the vertical geometric confinement as the free-slip boundaries do, but they also require the horizontal velocity to be zero at the two plates. Thus in the no-slip case the recirculation near the boundary within each convection cell is weakened, which in turn causes that the salt fingers are not as strong as in the free-slip case. 

\begin{figure*}
\begin{center}
\noindent\includegraphics[width=\textwidth]{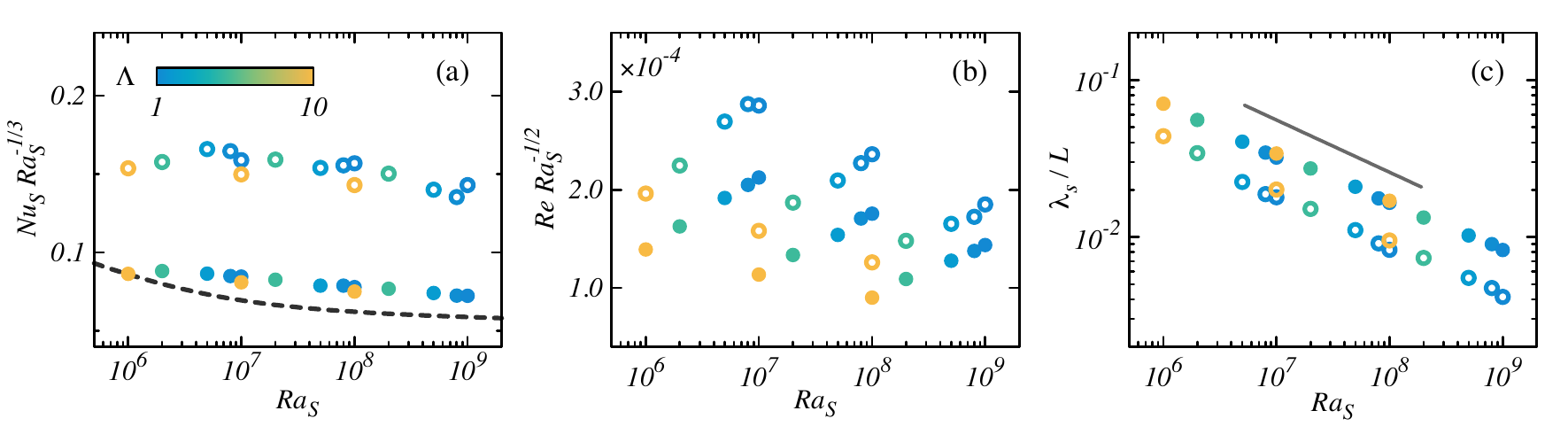}%
\caption{System responses versus the salinity Rayleigh number $Ra_S$. (a) Salinity Nusselt number $Nu_S$ compensated by $Ra_S^{1/3}$. (b) Reynolds number $Re$ compensated by $Ra_S^{1/2}$, and (c) the thickness of the salinity boundary layer $\lambda_s/L$ measured by the distance of the location of the first $s_{rms}$ peak and the adjacent boundary. The free-slip cases are marked by open symbols and the no-slip cases by solid symbols, respectively. Symbols are coloured according to the density ratio $\Lambda$. The dashed line in (a) corresponds to the Grossmann-Lohse prediction. The solid line in (c) has a slope of $-1/3$.}
\label{fig:resp}
\end{center}
\end{figure*}

The two most important responses of the system are the dimensionless salinity transfer rate $Nu_S$ and the dimensionless rms velocity $Re$, which are defined, respectively, as
\begin{equation}
Nu_S = \frac{\langle u_3 s \rangle - \lambda_S \partial_3 \langle s \rangle}{\lambda_S \Delta_S L^{-1}}, 
\quad Re = \frac{U_{rms}L}{\nu}.
\end{equation}
Here $\langle \cdot \rangle$ denotes the average over entire domain and time. The Reynolds number $Re$ is defined based on the rms value of velocity magnitude $U_{rms}$. The dependences of the salinity Nusselt number $Nu_S$ and the Reynolds number $Re$ as functions of $Ra_S$ are plotted in Fig.~\ref{fig:resp}. In Fig.~\ref{fig:resp}(a) we also compared the numerical results with the Grossmann-Lohse (GL) theory~\citep{GL2000,GL2001,GL2002,GL2004,GLrefit2013} for the no-slip cases. The agreement is very good, which has been found also in our previous studies~\citep{ddcjfm2015,ddcpnas16}. The responses of $Nu_S$ and $Re$ to different boundary conditions are consistent with the flow fields shown in Fig.~\ref{fig:finger} in two aspects. First, both quantities are enhanced by replacing the no-slip boundary condition with the free-slip one, which is attributed to stronger salt fingers as shown in Fig.~\ref{fig:finger}(c) as compared to \ref{fig:finger}(b). The salt fingers in the free-slip cases move faster in the vertical direction and therefore transfer salinity more efficiently. For the parameters considered here, the increment of $Nu_S$ is around $90\%$ and that of $Re$ around $35\%$, respectively. Second, the two quantities follow very similar scaling laws for different boundary conditions, which reflects the fact that the flow morphology is essentially the same, i.e. both cases are dominated by salt fingers. For both boundary conditions, $Nu_S$ follow the same trend. Our previous studies with no-slip boundary conditions revealed that for fixed $Ra_S$, as $\Lambda$ increases within the finger regime, $Nu_S$ only changes slightly and $Re$ decreases significantly~\citep{ddcjfm2015}. The present results suggest that it is also true for the free-slip cases.

We also compare the thickness $\lambda_s$ of the salinity boundary layer between the two boundary conditions in Fig.~\ref{fig:resp}(c). $\lambda_s$ is defined as the distance between the first peak of $s_{rms}$ profile and the adjacent boundary. The free-slip cases have smaller $\lambda_s$ compared to the no-slip cases, indicating that the salt fingers can reach much closer to the free-slip boundaries. Nevertheless, $\lambda_s$ follows the same power-law scaling with an exponent very close to $-1/3$, as indicated by the solid line in~\ref{fig:resp}(c).

\begin{figure*}
\begin{center}
\noindent\includegraphics[width=\textwidth]{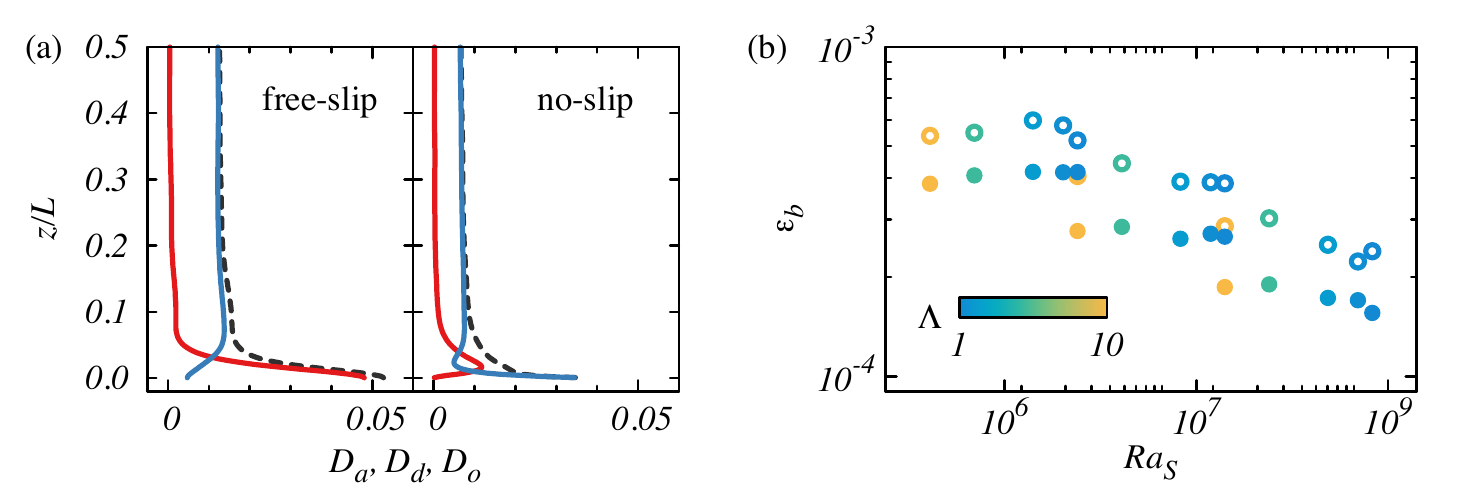}%
\caption{(a) The mean profiles of $D_a = \langle S_{ij}S_{ij} \rangle_{h}$ (black dashed lines) and the contributions from the diagonal components of the strain rate tensor $ D_d = \langle \Sigma_{i=j} (S_{ij}S_{ij}) \rangle_{h}$ (red lines) and the off-diagonal components $ D_o = \langle \Sigma_{i \neq j} (S_{ij}S_{ij}) \rangle_{h}$ (blue lines) for the free-slip case and no-slip cases. The control parameters are $Ra_S=5\times10^7$ and $\Lambda=2.0$. Thanks to the symmetry about the mid-height plane $z=0.5$, only the lower half of the domain can be shown. (b) The averaged dissipation rate $\varepsilon_b$ at boundary. The symbols and colormap are the same as in Fig.~\ref{fig:resp}.}
\label{fig:dsp}
\end{center}
\end{figure*}

The above observations indicate that salt fingers are robust with respect to different velocity boundary conditions, and therefore the exponents of the scaling laws for $Nu_S$ and $Re$ are also the same. According to the global balance between the Nusselt numbers and the total dissipation~\cite{ddcjfm2015}, the global dissipation rate must be higher for the free-slip cases than that for the no-slip cases. In Fig.~\ref{fig:dsp}(a) we plot the horizontally averaged mean-profiles of $D_a = \langle S_{ij}S_{ij} \rangle_h$ calculated from the flow fields shown in Fig.~\ref{fig:finger}. Here $S_{ij}=(\partial_j u_i + \partial_i u_j)/2$ is the strain-rate-tensor and it relates to the local dissipation rate by $\varepsilon=2 \nu S_{ij}S_{ij}$. Indeed, the dissipation rate of the free-slip case is higher in the bulk region than that of the no-slip case. Meanwhile, near the boundary the dissipation rate is also enhanced in the free-slip case. This is surprising since by imposing the free-slip boundary condition we eliminate the shear stress at the boundary. The increase in the dissipation rate near the boundaries when the no-slip boundary condition is replaced by the free-slip one is consistently observed for all Rayleigh numbers we simulated in the present study, as shown Fig.~\ref{fig:dsp}(b). Previous studies revealed similar behaviours in RB flow with no-slip or free-slip boundary conditions~\citep{Petschel2013}. 

To reveal the origin of the high dissipation rate, we divide $S_{ij}S_{ij}$ into the contribution from the diagonal components of $S_{ij}$, i.e. $D_d \equiv \langle \Sigma_{i=j} S_{ij}S_{ij} \rangle_h$, and that from the off-diagonal components $D_o \equiv \langle \Sigma_{i \neq j} (S_{ij}S_{ij}) \rangle_{h}$. The mean profiles of the two parts are also plotted in Fig.~\ref{fig:dsp}(a). In the bulk region, the dissipation for both boundary conditions is dominated by the off-diagonal components $D_o$. Further examination of the data reveals that the largest contribution is from the horizontal gradient of the vertical velocity, which corresponds to the shear among salt fingers as they move in the vertical direction. The free-slip case allows stronger fingers with larger vertical velocity, thus the dissipation rate is also higher than for the no-slip case at the same control parameters. 

\begin{figure*}
\begin{center}
\noindent\includegraphics[width=\textwidth]{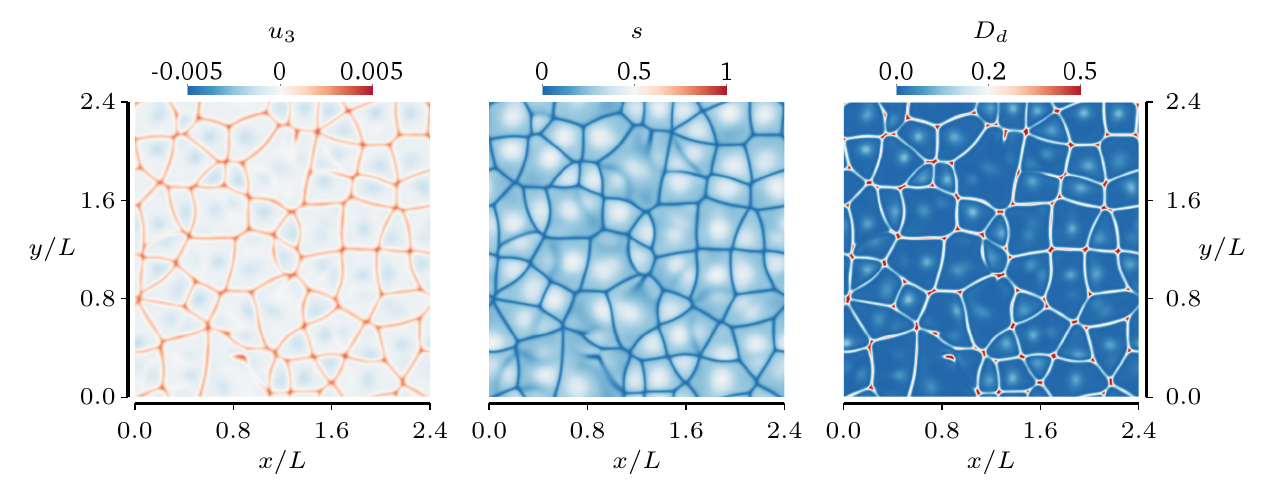}%
\caption{The contours on a horizontal plane near the bottom plate at $z/L=0.005$ for the flow field shown in Fig.~\ref{fig:finger}(c) with the free-slip boundary condition. Left: the vertical velocity $u_3$; middle: the salinity $s$; right: $D_d \equiv \Sigma_{i=j} (S_{ij}S_{ij})$ which dominates the local dissipation rate.}
\label{fig:zslice}
\end{center}
\end{figure*}

Near the boundary the situation is totally different. For the free-slip case the dissipation near the boundaries is dominated by $D_d$, as shown in Fig.~\ref{fig:dsp}(a) for $Ra_S=5\times10^7$ and $\Lambda=2.0$. This strong dissipation can be directly connected to the flow structures near two plates. In Fig.~\ref{fig:zslice} we show, for the free-slip case with the same parameters, the contours of vertical velocity $u_3$, salinity $s$, and the dominant terms of dissipation rate $D_d$ on a horizontal plane $z/L=0.005$ which is very close to the bottom plate. At this height the local dissipation is dominated by tangential shear stress in the no-slip case. The sheet structures are very distinct in the contours of both $u_3$ and $s$. These sheet structures rise from the bottom plate and carry a large salinity anomaly. The ascending fingers usually grow from the intersections of the sheet structures. When these structures move upward, they induce strong converging flows in the horizontal directions and therefore large dissipation occurs as shown in Fig.~\ref{fig:zslice}(c). Note that the locations with high dissipation rates coincide with the sheet structures observed in Figs.~\ref{fig:zslice}(a) and (b). Meanwhile, the descending fingers from the top plates decelerate as they reach the bottom plate and drive the expanding flow in the horizontal directions within the convection cells separated by the sheet structures, which corresponds to the large negative $u_3$ inside the cells in Fig.~\ref{fig:zslice}(a) and high salinity at the same locations in Fig.~\ref{fig:zslice}(b). Fig~\ref{fig:zslice}(c) indicates that at these locations the dissipation rate is also large.

At the no-slip boundary the velocity must be zero. In the thin layer adjacent to the plate, as shown in Fig.~\ref{fig:dsp}(a), the dissipation mainly originates from $D_o$ and the contribution from the diagonal components decreases to zero on the boundary. This is similar to a boundary layer where the vertical gradient of the horizontal velocity is the major source of dissipation. Above this quasi-boundary layer region, however, there exists a region where $D_d$ has a peak and it is larger than $D_o$. This is where the sheet structures and fingers start to grow and the mechanism of large dissipation is essentially the same as the high dissipation rate near the free-slip boundary. 

In conclusion, our study reveals that the vertically bounded fingering DDC is quanlitatively insensitive to different velocity boundary conditions, i.e. the no-slip or free-slip types, in the sense that the scaling laws keep the same for system responses such as the salinity Nusselt number and Reynolds number. Also for both boundary conditions the flow structures exhibit similar morphology. The free-slip boundary conditions allow stronger salt fingers which results in higher salinity transfer and flow velocity. However, even the tangential shear stress is eliminated at the boundaries by applying the free-slip conditions, the local dissipation rate is higher than that with the no-slip conditions in a thin layer adjacent to the boundaries. The strong dissipation is directly related to the vertically moving fingers and sheet structures near the boundary.

Two important indications can be derived based on the current study. First, the experiments for the DDC flows bounded by no-slip walls can still provide useful informations, especially for the scaling laws and salt fingers in the bulk region. Only the prefactors should be reevaluated for oceanic salt-finger layers. Second, the high dissipation rate near the free-slip boundaries suggests that large local dissipation is also likely to exist at interior interfaces, such as the boundaries of oceanic salt-finger layers. 

~

This study is supported by FOM and the National Computing Facilities, both sponsored by NWO, the Netherlands. The simulations were conducted on the Dutch supercomputer Cartesius at SURFsara, and the French supercomputer Curie through a PRACE project.

\end{document}